\documentclass[twocolumn,showpacs,aps,prl,superscriptaddress,preprintnumbers,amsmath,amssymb,letter,floatfix]{revtex4}

\usepackage{colordvi}
\usepackage{graphicx}
\usepackage{dcolumn}
\usepackage{bm}
\usepackage{multirow}

\input babarsym
\def\Rdcs       {\ensuremath{R_{\rm D}}\xspace} 

\def\Rws        {\ensuremath{R_{\rm WS}}\xspace}
\def\Rm         {\ensuremath{R_{\rm M}}\xspace}
\def\RmPm       {\ensuremath{R_{\rm M}^{\pm}}\xspace}
\def\RmP        {\ensuremath{R_{\rm M}^{+}}\xspace}
\def\RmM        {\ensuremath{R_{\rm M}^{-}}\xspace}
\def\xPrimeSq   {\ensuremath{{x^{\prime}}^2}\xspace}
\def\yPrimeSq   {\ensuremath{{y^{\prime}}^2}\xspace}
\def\xPrime     {\ensuremath{x^{\prime}}\xspace}
\def\xPrimeP    {\ensuremath{x^{\prime+}}\xspace}
\def\xPrimeM    {\ensuremath{x^{\prime-}}\xspace}
\def\yPrime     {\ensuremath{y^{\prime}}\xspace}
\def\yPrimeP    {\ensuremath{y^{\prime+}}\xspace}
\def\yPrimeM    {\ensuremath{y^{\prime-}}\xspace}
\def\xPrimePmSq {\ensuremath{{x^{\prime\pm}}^2}\xspace}
\def\xPrimeParenPmSq {\ensuremath{{x^{\prime(\pm)}}^2}\xspace}
\def\yPrimePmSq {\ensuremath{{y^{\prime\pm}}^2}\xspace}

\def\yPrimeParenPm   {\ensuremath{y^{\prime(\pm)}}\xspace}
\def\xPrimePSq  {\ensuremath{{x^{\prime+}}^2}\xspace}
\def\xPrimeMSq  {\ensuremath{{x^{\prime-}}^2}\xspace}
\def\AD         {\ensuremath{A_{\rm D}}\xspace}
\def\AM         {\ensuremath{A_{\rm M}}\xspace}

\def\Dztokpi    {\ensuremath{\Dz \to K^{-}\pi^{+}}\xspace}

\def\DztokpiWS  {\ensuremath{\Dz \to K^{+}\pi^{-}}\xspace}

\def\mKpi       {\ensuremath{m_{K\pi}}\xspace}

\def\dm         {\ensuremath{\delta m}\xspace}

\newcommand{\kevcc}{\ensuremath{{\mathrm{\,Ke\kern -0.1em V\!/}c^2}}\xspace}

\newcommand{\BABARPubYear}    {03}
\newcommand{\BABARPubNumber}  {009}

\newcommand{\SLACPubNumber} {9685}

\iffalse 
\def\deleted{\skip0=0.5\lastskip\unskip\hskip\skip0\relax
  \hbox to0pt{\hss\Blue{\vrule height 2ex depth 0.3ex}\hss}\hskip\skip0\relax}

\else
\let\deleted=\relax

\fi

\long\def\inst#1{\par\nobreak\kern 4pt\nobreak
    {\it #1}\par\vskip 10pt plus 3pt minus 3pt}

\begin{document}

\preprint{\babar-PUB-\BABARPubYear/\BABARPubNumber} 
\preprint{SLAC-PUB-\SLACPubNumber}

\title{Search for {\boldmath $\Dz$}-{\boldmath $\Dzb$} Mixing
  and a Measurement of the Doubly Cabibbo-suppressed Decay
  Rate in
  {\boldmath $\Dz\ra K\pi$}~Decays}

%
%
\author{B.~Aubert}
\author{R.~Barate}
\author{D.~Boutigny}
\author{J.-M.~Gaillard}
\author{A.~Hicheur}
\author{Y.~Karyotakis}
\author{J.~P.~Lees}
\author{P.~Robbe}
\author{V.~Tisserand}
\author{A.~Zghiche}
\affiliation{Laboratoire de Physique des Particules, F-74941 Annecy-le-Vieux, France }
% ...
\collaboration{The \babar\ Collaboration}
\noaffiliation

\date{\today}

\begin{abstract}
  We present results of a search for \Dz-\Dzb~mixing and a
  measurement of $\Rdcs$, the ratio of doubly
    Cabibbo-suppressed decays to Cabibbo-favored decays, based on an
  analysis of \DztokpiWS decays in 57.1~\invfb of data collected at or
  just below the \Y4S resonance with the \babar\ detector at the \pep2
  collider.  Our results are compatible with no mixing and no \CP
  violation.  At the 95\% confidence level, allowing for \CP violation,
  we find the mixing parameters $\xPrimeSq<0.0022$
  and $-0.056<\yPrime<0.039$, and the mixing rate $\Rm <
  0.16\%$.  In the limit of no mixing, \deleted $\Rdcs = ( 0.357 \pm
  0.022 \hbox{ (stat.)} \pm 0.027 \hbox{ (syst.)})\%$ and the
  \CP-violating asymmetry \mbox{\deleted$\AD = 0.095 \pm 0.061 \hbox{
      (stat.)} \pm 0.083 \hbox{ (syst.)}$\deleted.}
\end{abstract}

\maketitle

Within the Standard Model the level of \Dz-\Dzb~mixing is predicted to
be below the sensitivity of current
experiments~\cite{Falk:2001hx}. For this reason \Dz-\Dzb~mixing is
a good place to look for signals of new physics beyond the 
Standard Model~\cite{Nelson:1999fg}.
Because new physics may not conserve \CP, it
is important to consider \CP violation when measuring mixing. Observation
of \CP violation 
in \Dz-\Dzb~mixing
would be an unambiguous sign of new
physics~\cite{Falk:2001hx,Blaylock:1995ay}.

Mixing can be characterized by the two parameters $x \equiv \Delta m /
\Gamma$ and $y \equiv \Delta\Gamma/2\Gamma$, where $\Delta m = m_1 - m_2$
($\Delta\Gamma = \Gamma_1 - \Gamma_2$) 
is the difference in mass (width) between the two
mass eigenstates and $\Gamma$ is the average width. 

The dominant two-body decay of the \Dz is the
\emph{right-sign} (RS) Cabibbo-favored~(CF) decay \Dztokpi. 
Evidence for mixing and \CP violation, if present, will 
appear in the 
\emph{wrong-sign}~(WS) decay~\DztokpiWS.  Charge conjugates are implied
unless otherwise stated. 
Two amplitudes contribute to the production of this final
state:
the tree-level amplitude for
doubly Cabibbo-suppressed~(DCS) decay of the \Dz,
and an amplitude for mixing followed by CF~decay of the \Dzb.
Assuming 
that $x$,~$y \ll 1$ and \CP 
is conserved, and with the convention
$\Delta\Gamma=\Gamma(\CP=+1)-\Gamma(\CP=-1)$,
the
time-dependent, WS decay rate~$T_{\rm WS}(t)$ for $\Dz\ra\Kp\pim$
can be approximately~\cite{Godang:1999yd}
related to the RS~decay rate~$T_{\rm RS}(t)$ by
\begin{equation}
    \label{eq:TimeEvolNoCPV}
    T_{\rm WS}(t) = T_{\rm RS}(t) 
    \left( 
      \Rdcs + 
      \sqrt{\Rdcs}\yPrime\; t + 
      \frac{\xPrimeSq + \yPrimeSq}{4} t^2
    \right).
\end{equation}
In 
Eq.~(\ref{eq:TimeEvolNoCPV}),
$t$ is the proper time of the 
\Dz~decay measured in units of the 
$\Dz$~lifetime~$\tau_{\Dz}$, $T_{\rm RS}(t)\propto e^{-t}$, 
$\Rdcs$ is the time-integrated rate of the direct 
DCS decay \DztokpiWS relative to the RS~decay,
and $\xPrime$,~$\yPrime$ are related to $x$,~$y$ by 
$\xPrime = x\cos\delta_{K\pi} + y \sin\delta_{K\pi}$ and 
$\yPrime = -x\sin\delta_{K\pi} + y \cos\delta_{K\pi}$,
where $\delta_{K\pi}$ is the relative strong phase 
between the CF and DCS amplitudes. Physics 
beyond the Standard Model may include additional phases
that are not \CP-conserving. Such terms
can be absorbed into a phase $\varphi$, described below.
The time-integrated WS decay rate is
\begin{equation}
\Rws = \Rdcs + \sqrt{\Rdcs}\yPrime + \frac{\xPrimeSq+\yPrimeSq}{2} \;.
\label{eq:integralNoCPV}
\end{equation}

Previous
experiments 
have searched for mixing 
using wrong-sign 
hadronic~\cite{Anjos:1988pw,Aitala:1998fg,Godang:1999yd} and 
semileptonic~\cite{Aitala:1999dt}
\Dz decays, or have searched
for width differences between $\CP=+1$ and $\CP=-1$ states 
directly~\cite{Link:2000cu,Csorna:2001ww,Abe:2001ed}.
Since $\xPrime$ appears only quadratically
in Eq.~(\ref{eq:TimeEvolNoCPV}), its sign cannot be determined 
in an analysis based on the WS~decay alone.


To allow for \CP violation, we apply Eq.~(\ref{eq:TimeEvolNoCPV}) 
to \Dz and \Dzb separately. We determine
\{$\Rws^+$, $\xPrimePSq$, $\yPrimeP$\} for \Dz candidates and 
\{$\Rws^-$, $\xPrimeMSq$, $\yPrimeM$\} for \Dzb candidates.
The separate \Dz and \Dzb results can be combined to form the quantities
\begin{equation}
  \label{eq:ADdef}
  \AD = \frac{\Rdcs^{+} - \Rdcs^{-}}{\Rdcs^{+} + \Rdcs^{-}}
        \hbox{,\qquad}
  \AM = \frac{\RmP - \RmM}{\RmP + \RmM} \;,
\end{equation}
where $\RmPm \equiv (\xPrimePmSq + \yPrimePmSq) /2$.
\AD and \AM are related to \CP violation in the DCS decay
and mixing amplitudes, respectively. \CP violation in the interference
of DCS decay and mixing is parameterized by the phase $\varphi$:
\begin{eqnarray}
  \label{eq:WS_CPx}
  {\xPrime}^\pm  & = & \sqrt[4]{\frac{1 \pm \AM}{1 \mp \AM}}
                ( \xPrime\cos\varphi \pm \yPrime\sin\varphi ), \\
  \label{eq:WS_CPy}
  {\yPrime}^\pm  & = & \sqrt[4]{\frac{1 \pm \AM}{1 \mp \AM}} 
                ( \yPrime\cos\varphi \mp \xPrime\sin\varphi ).
\end{eqnarray}
An offset in $\varphi$ of $\pm\pi$ can be absorbed
by a change in sign of both $\xPrime$ and $\yPrime$, 
effectively swapping the definition of
the two physical \Dz states without any other observable consequence.
To avoid this ambiguity, we use the convention that $|\varphi| < \pi/2$.

We select a very clean sample of RS~and WS~decays and 
fit for signal and background components in
a $57.1~\fb^{-1}$ dataset collected with the 
\babar\ detector~\cite{Aubert:2001tu}
at the PEP-II $e^+e^-$ storage ring.  We extract the parameters 
describing mixing and DCS amplitudes from the WS~decay-time distribution.
To avoid potential bias,
we finalized our data selection criteria and the procedures for 
fitting and extracting the statistical limits 
without
examining the mixing results.

We select \Dz candidates from reconstructed $\Dstarp \ra \Dz\pip$
decays; this provides a clean sample of \Dz decays, and the
charge of the pion (the `tagging pion')
identifies the production flavor of the neutral~$D$.
We retain each 
RS and WS \Dz candidate
whose invariant
mass~\mKpi is within $ 60$~\mevcc of the 
\Dz~mass.
We require the mass difference~\dm between the~\Dstarp and the~\Dz candidate
to be less than $m_{\pi}+25$~\mevcc.
Only
\Dstarp candidates with center-of-mass momenta above $2.6$~\gevc are
retained, thereby rejecting
\Dstarp candidates from $B$~decays.

We determine the \Dz vertex by requiring that the \Dz~decay tracks
originate from a common point with a probability $p(\chi^2) > 1\%$, and
then determine the 
\Dstarp vertex by extrapolating the \Dz flight path back to
the beam-beam interaction region. This procedure benefits from the small
vertical size ($\approx 7$~$\mu$m) of the luminous region and the
well-measured \Dz decay products. We 
constrain
the 
trajectory of the tagging pion 
to originate from the \Dstarp vertex, 
thus improving the measurement of \dm.
We then calculate the proper time~$t$ of the \Dz~decay
from the dot product of the \Dz momentum vector and flight vector, 
defined by the \Dstarp and \Dz decay vertices in three dimensions.
The typical resolution is 0.2~ps. 

We determine the mixing parameters by unbinned, 
extended maximum-likelihood fits to the RS and WS samples
simultaneously.
We perform four separate fit cases:
first, a general fit allowing for possible \CP~violation,
which treats WS \Dz and \Dzb candidates
separately, fitting for
\{$\Rws^+$, $\xPrimePSq$, $\yPrimeP$\} for \Dz candidates and 
\{$\Rws^-$, $\xPrimeMSq$, $\yPrimeM$\} for \Dzb candidates; second, a fit 
assuming \CP~conservation, which does not differentiate between \Dz and \Dzb 
candidates, fitting for \{$\Rws$, $\xPrimeSq\!\!$, $\yPrime$\};
third, a fit assuming no mixing, but allowing
\CP violation in the DCS amplitudes,
fitting for \{\Rdcs, $A_{\rm D}$\}; and fourth,
a fit for \Rdcs, only, assuming \CP conservation and no mixing.

For each fit case we assign each candidate to one of four categories based on
its origin as \Dz or \Dzb, and its decay as RS or WS.
For each category we construct probability density functions~(PDFs)
that model signal and background components.
As independent input variables in the PDFs we use
the \Dz candidate
mass~\mKpi, the mass difference~\dm,
and the \Dz proper time~$t$ with its error~$\sigma_t$.

The fit is performed by simultaneously maximizing
individual extended likelihood functions,
one for each candidate category. 
Within each category, the likelihood is a sum of PDFs, one for each
signal or background component, weighted by the 
number of events
for that component.
Each component's PDF factorizes into a portion
describing the behavior of each independent variable convoluted with a
corresponding resolution function.
The parameters describing the mass resolutions and shapes and the
lifetime resolution are shared between PDFs.
These parameters are determined primarily by the much
larger RS~sample.

We characterize the WS background by three components: true
\Dz decays that are combined with unassociated pions to
form \Dstarp candidates; 
combinatorial background where
one or both of the tracks in the \Dz candidate do not originate from
\Dz decay; 
and background where the kaon and the pion in a \Dz
decay have both been misidentified, thus converting a RS decay
into an apparent WS decay (double misidentification).  
Kaons (pions) are identified
with an average efficiency of 84\%~(85\%); 
the average misidentification rate is
3\%~(2\%).
Correctly fitting the WS~double misidentified background is particularly 
important due to the large size of the RS~sample; its level as 
obtained from the fit agrees well with predictions based on our
particle~identification performance.

We treat the 
normalization of WS candidates originating as a \Dz or \Dzb
separately, thus yielding in total two signal and six
background components in the WS part of the PDF. 
We assume \CP~conservation in the RS data; its PDF
has one signal and three background components.  

\begin{figure}[htb]
  \centering
  \includegraphics[width=0.99\linewidth, clip=]{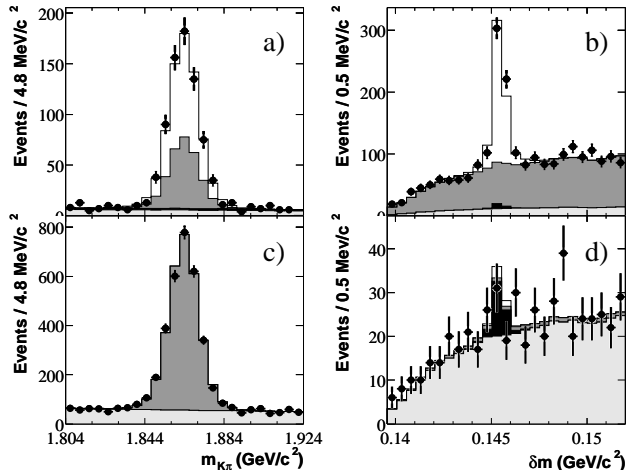}
  \caption{The distribution of the WS data for 
    a) \mKpi with $144.5<\dm<146.5~\mevcc$, 
    b) \dm with $|\mKpi-m_{\Dz}|<20~\mevcc$, 
    c) \mKpi with $150<\dm<165~\mevcc$,
    and d) \dm with $25<|\mKpi-m_{\Dz}|<60~\mevcc$. 
    Data are shown as points with the 
    contributions from the fit overlaid:
    signal (open), unassociated pion background (dark shaded), double
    misidentification
    background (black), and combinatorial background (light shaded).}
  \label{fig:WSmass}
\end{figure}

We perform each fit 
in steps. 
Parameters corresponding to the \mKpi and \dm
distributions and the number of candidates in each category are
determined first.  In a second step, these parameters are fixed
and a fit to the proper time distribution is performed.
The shapes of the distributions 
in \mKpi and \dm allow the fit to differentiate between the
various signal and background components.
Figure~\ref{fig:WSmass} 
shows projections from the WS sample 
of the \mKpi and \dm distributions overlaid on
the fit result.

We fit the RS decay-time distribution using a model
that combines the
RS signal decay-time distribution
($T_{\rm RS}(t)$ in Eq.~(\ref{eq:TimeEvolNoCPV}))
and the expected decay-time distributions of each background
component,
convolving each with a common decay-time resolution model
that uses the 
decay-time error for each
candidate and a scaling factor determined in the fit.
For the WS signal component we use the same resolution 
model 
but with a lifetime distribution including the mixing parameters as
given by $T_{\rm WS}(t)$ in Eq.~(\ref{eq:TimeEvolNoCPV}) or its \CP-violating
counterparts.
For the unassociated pion and double
misidentification backgrounds we also use the 
$T_{\rm RS}(t)$~lifetime distribution because they are
true \Dz decays. The
combinatorial background is assigned a zero-lifetime distribution 
and a signal-type resolution model based on
studies of mass sidebands and Monte Carlo~(MC) samples.

\begin{table}[htb]
  \centering
  \caption{Fit parameter results determined by the full fit, 
           with no constraint on $\xPrimeSq$ 
           in the mixing-allowed cases.
           For the no-mixing cases, $\Rws^{(\pm)}=\Rdcs^{(\pm)}$.
           The $+$($-$) signifies \Dz(\Dzb).}
  \vskip 0.1 in
  \begin{tabular}{lclclcl}
  \hline
     Fit case & Parameter & \multicolumn{5}{c}{Fit result (${}/10^{-3}$)} \\
              &     &  \multicolumn{1}{c}{\hspace{3mm}\Dz} 
 & \hspace{3mm} & \multicolumn{1}{c}{\hspace{3mm}\Dzb}
 & \hspace{7mm} & \multicolumn{1}{c}{$\Dz + \Dzb$}\\
  \hline
 \multirow{3}{1.5cm}{Mixing \\ allowed}
      & $\Rws^{(\pm)}$      & $\phantom{-0}3.9$   && $\phantom{-0}3.2$ && $\phantom{-}3.6$\\
      & $\xPrimeParenPmSq$ & $\phantom{0}$\relax$-0.79$  &&
     $\phantom{0}$\relax$-0.17$ && $-0.32$ \\
      & $\yPrimeParenPm$   & $\phantom{-}17$    && $\phantom{-}12$ && $\phantom{-}13$ \\
    \hline
 No mixing
  & $\Rws^{(\pm)}$ & $\phantom{-0}3.9$ && $\phantom{-0}3.2$ && $\phantom{-}3.6$\\ 
    \hline
  \end{tabular}
  \label{tab:fitresults}
\end{table}

\begin{figure}[htb]
  \centering
  \includegraphics[width=0.99\linewidth, clip=]{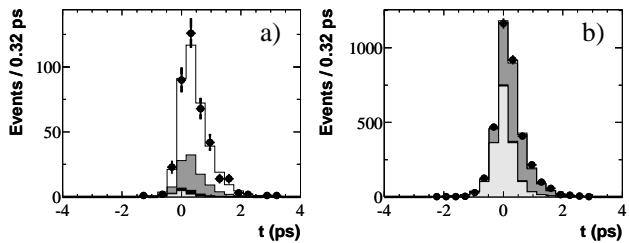}
  \caption{The proper time distribution for the WS candidates in a) the signal
    region (73\% signal purity) and b) a background region (50\% combinatorial
    background). 
    See Fig.~\protect\ref{fig:WSmass} for component definitions.}

  \label{fig:WStime}
\end{figure}

In Table~\ref{tab:fitresults} we summarize the central values
returned by the fit for the four cases.
In Fig.~\ref{fig:WStime} 
we show the decay-time distribution of the WS
sample
for the signal
and a background region.
We select a signal (background) region to provide a sample
with 73\% signal (50\% combinatorial background) candidates based on
the reconstructed values of \mKpi and \dm. The selected signal region
contains 64\% of all signal events according to the fit. In total we
observe about 120,000~RS (430~WS) signal decays.

Our fit 
permits
$\xPrimeSq$ to take unphysical negative values.
The interpretation of non-physical results and error estimates
calculated from the 
log-likelihood surface~(LLS) would require a
Bayesian analysis where the choice of prior is not clear.
In addition,
an accurate error estimate from the LLS requires a LLS shape that is
not strongly dependent on the outcome of the fit. 
These requirements are not satisfied here.
Therefore, we use a frequentist approach,
and construct
95\% confidence-level (CL) contours in 
$(\xPrimeSq\!\!,\yPrime)$ 
utilizing toy~MC experiments.
In each toy~MC experiment we
generate a WS~dataset (the part sensitive to mixing)
for a given $(\xPrimeSq\!\!,\yPrime)$
with 
the same number of \Dz and \Dzb events as observed in the
data, but with a decay-time distribution appropriate for the
chosen point.
Fit parameters for the \mKpi and \dm 
distributions and other parameters not sensitive to mixing are
fixed at their fitted values from data.  
The $\sigma_t$
distribution and background fractions from the data fit are used as well.
We fit each toy~MC dataset, obtaining values 
for the mixing parameters and the corresponding LLS.
We construct contours such that for any point
$\vec\alpha_c=(\xPrimeSq_c, \yPrime_c)$ on the contour
95\% of the experiments 
generated at that point will have a log-likelihood difference
$\Delta \ln {\cal L}(\vec\alpha_c)
= \ln {\cal L}_{\rm max} - \ln {\cal L}(\vec\alpha_c)$
less than the corresponding value $\Delta \ln {\cal L}_{\rm
  data}(\vec\alpha_c)$ evaluated for the data. ${\cal L}_{\rm max}$
is the maximum likelihood obtained from a fit to either data or
a toy MC~sample. 

\begin{figure}[htb]
  \includegraphics[width=0.95\linewidth]{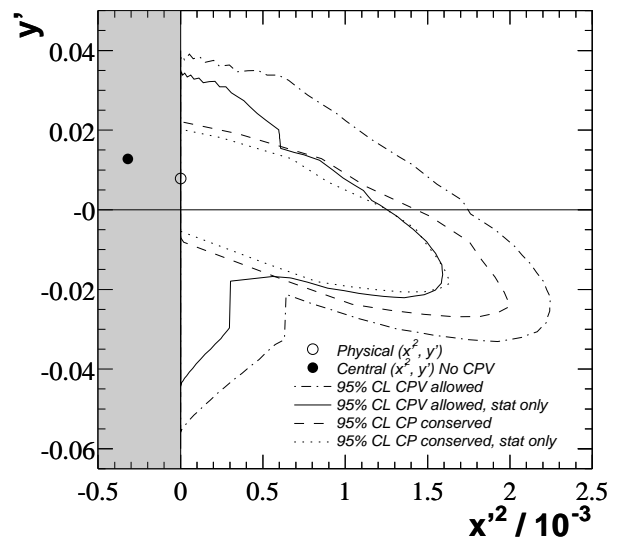}
  \caption{95\% CL limits
    in $(\xPrimeSq\!\!, \yPrime)$ with and
    without \CP violation (CPV) allowed. The solid point represents the most
      likely fit point assuming \CP conservation
      and the open circle the same but allowing \CP violation and forcing
      $\xPrimeSq>0$. The dotted (dashed) line is the statistical
      (statistical and systematic) contour for the case where no \CP
      violation is allowed. The solid and dash-dotted lines are 
      for the corresponding case where \CP violation is allowed.}
  \label{fig:Contour}
\end{figure}

Where we assume \CP conservation we apply this method to the
combined \Dz and \Dzb WS samples.
The resulting contour is
shown by the dotted line in Fig.~\ref{fig:Contour}.
The 95\%~CL limits for \Rdcs and for \Rm are obtained by
finding their extreme values on the 95\%~CL contour.

To consider \CP violation, we divide the
WS sample into candidates produced as a \Dz or as a \Dzb and
calculate separate contours for $(\xPrimePSq,
\yPrimeP)$ and $(\xPrimeMSq, \yPrimeM)$, each
corresponding to a CL of $1-\sqrt{0.05} = 77.6\%$. Each
point on the \Dz contour is combined with each point on the \Dzb
contour using Eqs.~(\ref{eq:ADdef})--(\ref{eq:WS_CPy})
to produce two potential solutions of
\mbox{\{$\xPrimeSq\!\!$, $\yPrime$\}} for each relative sign 
of $\xPrimeP$ and $\xPrimeM$.
The outer envelope of these
points is presented as the 95\% CL contour in the 
$(\xPrimeSq\!\!,\yPrime)$ plane (see Fig.~\ref{fig:Contour}).
The peculiar shape of the contour arises from 
the two potential solutions for each point on the \Dz and \Dzb contours.
This contour is more stringent than the \CP-conserving case in some cases,
which is allowed as the definition of coverage is
slightly different. 
No central value for \xPrimeSq exists 
if either \xPrimePSq or $\xPrimeMSq < 0$.

We summarize results for all four fit cases
in Table~\ref{tab:results}. We obtain limits on the
individual mixing parameters by projecting the contours onto the
corresponding coordinate axes. Since the no-mixing solution is well within
the 95\%~CL contour, we cannot place limits on \AM and~$\varphi$.

\begin{table}[htb]
  \centering
  \caption{A summary of our results including systematic errors.
      A central value is reported for the full fit with $\xPrimeSq$
      fixed at zero. The 95\% CL limits are for the case where
      $\xPrimeSq$ was not constrained during the fit.}
  \vskip 0.1 in
  \begin{tabular}{lccr@{$\;<\;$}l}
    \hline
     Fit case & Parameter & \multicolumn{1}{c}{Central value} &  
     \multicolumn{2}{c}{95\% CL interval} \\
     &   & ($\xPrimeSq\!\!=\!0$) (${}/10^{-3}$) & 
     \multicolumn{2}{c}{(${}/10^{-3}$)} \\
    \hline
 \multirow{4}{1.7cm}{\CP\\ violation \\ allowed}
      & $\Rdcs$     & $3.1$  & $2.3 $&$ \Rdcs < 5.2$ \\
      & $\AD$       & $1.2$  & $-2.8 $&$ \AD < 4.9$ \\
      & $\xPrimeSq$ & $0\phantom{.0}$  & $\xPrimeSq $&$ 2.2$ \\
      & $\yPrime$   & $8.0$ & $-56 $&$ \yPrime<39$ \\
      & $\Rm$       &       & $\Rm$ & $1.6$ \\
    \hline
 \multirow{3}{1.7cm}{No \CP\\ violation}
  &  $\Rdcs$      & $3.1$  & $2.4 $&$ \Rdcs < 4.9$ \\
  &  $\xPrimeSq$  & $0\phantom{.0}$     & $\xPrimeSq $&$ 2.0$  \\
  &  $\yPrime$    & $8.0$ & $-27 $&$\yPrime<22$ \\
      & $\Rm$       &       & $\Rm$ & $1.3$ \\
    \hline
 \multirow{2}{1.7cm}{No mixing}
  & \multicolumn{4}{l}{\,\,\,$\Rdcs = ( 0.357 \pm 0.022 \hbox{ (stat.)} 
      \pm 0.027 \hbox{ (syst.)})\%$} \\
  & \multicolumn{4}{l}{\,\,\,$\AD =  0.095 \pm 0.061 \hbox{ (stat.)} 
      \pm 0.083 \hbox{ (syst.)}$} \\
    \hline
No \CP viol. & \multicolumn{4}{l}{\multirow{2}{6.4cm}{\,\,\,$\Rdcs = ( 0.359 \pm 0.020 \hbox{ (stat.)} 
      \pm 0.027 \hbox{ (syst.)})\%$}} \\
or mixing \\
    \hline
  \end{tabular}
  \label{tab:results}
\end{table}
To estimate systematic uncertainties we evaluate the contributions from
uncertainties in the parametrization of the PDFs, detector effects,
and event selection criteria.
The small systematic 
effects of fixing the \mKpi and \dm parameters and the number of events 
in each category in the final fit is evaluated by varying 
these parameters within statistical uncertainties
while accounting for statistical
correlations.

For detector effects such as alignment errors or charge asymmetries we
measure their effect on the RS~sample. 
Under the assumption that 
RS decay is exponential and has no direct \CP violation, this method is
very sensitive.
The systematic error due to the size of the MC~sample is insignificant
since all distributions are obtained from the data.

Each systematic check yields a small shift in the fitted mixing
parameters. We use MC~experiments to determine the
significance of each shift using the same method employed as
for the 95\%~CL statistical contour. 
We scale the statistical contour 
with respect to the central fitted point by the factor $\sqrt{1
  + \sum m_i^2}$, where $m_i$ is the relative significance of each
systematic check. For the general case we carry out this procedure 
for the \Dz and \Dzb contours separately before combination. 
In all fits the largest effect for \xPrimeSq and \yPrime is 
the \Dstarp momentum selection cut,
with $m_i^2=0.24$; all others are at least three times smaller. 
For \Rdcs the largest effect is the decay-time range.
We
show contours including systematic errors in
Fig.~\ref{fig:Contour} as a dashed line in the \CP conserving case and as
a dash-dotted line in the general case. 

In summary, we have set improved limits on  $ \Dz $-$ \Dzb $
mixing and on \CP
violation in WS decays of neutral $D$ mesons. 
Our results are compatible with previous
measurements~\cite{Anjos:1988pw,Aitala:1998fg,Godang:1999yd} and with no
mixing and no \CP violation, 
which agrees with Standard Model predictions.

\begin{acknowledgments}
We are grateful for the excellent luminosity and machine conditions
provided by our \pep2\ colleagues, 
and for the substantial dedicated effort from
the computing organizations that support \babar.
The collaborating institutions wish to thank 
SLAC for its support and kind hospitality. 
This work is supported by
DOE
and NSF (USA),
NSERC (Canada),
IHEP (China),
CEA and
CNRS-IN2P3
(France),
BMBF and DFG
(Germany),
INFN (Italy),
FOM (The Netherlands),
NFR (Norway),
MIST (Russia), and
PPARC (United Kingdom). 
Individuals have received support from the 
A.~P.~Sloan Foundation, 
Research Corporation,
and Alexander von Humboldt Foundation.

\end{acknowledgments}

\bibliography{apssamp}

\end{document}